\documentstyle[12pt]{article}

\input epsf

\begin{document}
\begin{titlepage}
\today          \hfill
\begin{center}
\hfill    OITS-696 \\

{\large \bf
Remarks on models with singlet neutrino in large
extra dimensions
}
\footnote{This work is supported by DOE Grant DE-FG03-96ER40969.}
\vskip .15in
Kaustubh Agashe \footnote{email: agashe@oregon.uoregon.edu},
Guo-Hong Wu \footnote{email: wu@dirac.uoregon.edu}

{\em
Institute of Theoretical Science \\
University
of Oregon \\
Eugene OR 97403-5203}
\end{center}

\vskip .05in

\begin{abstract}
Small Dirac masses for neutrinos are natural in models with singlet 
fermions in large extra dimensions with
quantum gravity scale $M_{\ast} \sim 1- 100$ TeV.
We study two modifications of the minimal model in order to obtain the
mass scale relevant 
for atmospheric neutrino oscillations with at most $O(1)$
higher-dimensional
Yukawa 
couplings and with $M_{\ast} \sim$ a few TeV. 1) In models with
singlet fermions
in smaller number of extra dimensions than gravity, we
find that the effects on BR$ \left( \mu \rightarrow e \gamma \right)$ 
and on
charged-current universality in $\pi^- \rightarrow e
\bar{\nu}, \; \mu \bar{\nu}$ decays
are suppressed
as compared to that in
the minimal model with neutrino and gravity in the same
space. 2)
If small Dirac masses for the singlets are added
along with lepton number violating couplings, then the mass scales
and mixing angles for neutrino oscillations
can be different from those
relevant for $\mu \rightarrow e \gamma$
and $\pi^- \rightarrow e \bar{\nu}, \; \mu \bar{\nu}$.
Thus, in both modified models
the constraints on $M_{\ast}$ from 
BR$ \left( \mu \rightarrow e \gamma \right)$ and
$\pi^- \rightarrow e \bar{\nu}, \; \mu \bar{\nu}$
decays can be significantly relaxed.
Furthermore,
constraints from supernova $1987$a 
strongly disfavor oscillations of active
neutrinos to sterile neutrinos in both the minimal and the modified models.

\end{abstract}

\end{titlepage}

\newpage
\renewcommand{\thepage}{\arabic{page}}
\setcounter{page}{1}

{\bf \large 1) Introduction}

In models with large extra dimensions and low ($\sim$ TeV) quantum gravity
scale \cite{add}, 
it is well known that neutrino masses can be naturally
small if the singlet (right-handed) neutrino propagates in extra 
dimensions \cite{aetal, ddg, others}. 
In the simplest version of these models, if
the quantum gravity scale is
smaller than ${\sim}10$ TeV (as motivated by the hierarchy
problem) and if the higher-dimensional
Yukawa couplings are at most $O(1)$,
then it is difficult to obtain
the neutrino mass scale required for a solution to the atmospheric
neutrino anomaly. 
On the other hand, 
charged-current universality in
$\pi^- \rightarrow e \bar{\nu}, \; \mu \bar{\nu}$
decays and the limit on
BR$\left( \mu \rightarrow e \gamma \right)$ 
constrain
the quantum gravity scale to be larger than
$10-100$ TeV, almost independent of the number of
extra dimensions (assuming large $\nu _e -
\nu _{\mu}$ mixing).

In this paper,  
we discuss two modifications
of this model where it is possible to obtain
the mass scale required for atmospheric
neutrino oscillations even with a quantum gravity scale
of a few TeV
and Yukawa couplings of $O(1)$. Furthermore,
we will show that in these models, 
this quantum gravity scale can be consistent
with 
charged-current universality in 
$\pi ^- \rightarrow e \bar{\nu}, \; \mu \bar{\nu}$
decays and the limit on 
BR$\left( \mu \rightarrow e \gamma \right)$.

Consider for simplicity standard model (SM) gauge singlets $N_I = (\psi _I,
\bar{\chi}_I)$ in $5D$, where $\psi_I$, 
$\chi_I$ are $2$-component Weyl spinors. 
Assume the following terms in the $5D$ action:
\begin{eqnarray}
S_{\hbox{free}} & = & \int d^4 x dy \; i
\bar{N}_I \Gamma _{a} \partial _a N _I + \int d^4 x dy \left(
\mu_I \bar{N}_I N_I + \hbox{h.c.} \right) +
\int d^4 x \; i \bar{\nu}_f \sigma_{\mu} \partial _{\mu} \nu_f ,
\nonumber \\
S_{\hbox{int}} & = & 
\int d^4 x \left( \frac{\lambda _{fI}}{\sqrt{M_{\ast}}} 
\nu _f \psi _I (
y=0) \;
h + \frac{\lambda ^c _{fI}}{\sqrt{M_{\ast}}} \nu _f \chi _I
(y=0) \; h + \hbox{h.c.} \right),
\nonumber \\
 & & 
\end{eqnarray}
where $y$ denotes the extra dimension assumed to be compactified on
a circle of radius $R$ and $M_{\ast}$ is the $5D$
``fundamental''
Planck scale.
$\lambda$, $\lambda ^c$ are dimensionless Yukawa couplings
in $5D$. We have allowed Dirac mass terms ($\mu$)
for singlets and 
lepton number violating Yukawa couplings, $\lambda ^c$ (assigning
lepton numbers $+1$ and $-1$ to $\nu_f$ and $N_I$, respectively).
The index $\mu$ runs over $4D$ while $a$ runs over $5D$ and
$\Gamma _a$ are the ``gamma'' matrices in $5D$.
$f$ is
flavor index for SM neutrinos and $I$ denotes the index for the singlets
and $h$ is the Higgs doublet.
We have chosen the basis for $N_I$
in which the mass matrix $\mu$ is diagonal.

In the effective $4D$ theory, $\psi _I$ and $\chi _I$ appear as 
towers of Kaluza-Klein (KK) states, $\psi _I (x,y) = 
\sum _n 1/\sqrt{2 \pi R} \; e^{iny/ R} \psi^{(n)} _I (x)$ and 
%\\ $\chi _I (x,y) = 
similarly for $\chi _I (x,y)$
%\sum _n 1/\sqrt{2 \pi R} \; e^{iny/ R} \chi^{(n)}_I (x)$ 
giving
\begin{eqnarray}
S_{\hbox{mass, int}} 
& = & \int d^4x \sum _n
\left[ \left( \mu _I + \frac{in}{R} \right)
\psi _I ^{(n)} \chi _I ^{(n)} + \right. \nonumber \\
 & &  \left. \frac{ \lambda _{fI} }{ \sqrt{ 2 \pi } } 
\frac{M_{\ast}}{M_{Pl}} \nu _f \psi _I ^{(n)} h  +
\frac{ \lambda _{fI}
^c }{ \sqrt{ 2 \pi } }
\frac{M_{\ast}}{M_{Pl}} \nu _f \chi _I ^{(n)} h + \hbox{h.c.} \right].
\end{eqnarray}
Here we have used the relation, $M_{\ast}^{\delta + 2} R^{\delta}
\sim M_{Pl}^2$ 
(with $M_{Pl} \sim 2.4 \times 10^{18}$ GeV), where
$\delta$ is the number of extra spatial dimensions all of which are 
assumed to be of the same
size 
\footnote{Of course, $\delta = 1$ is not realistic since in that case
$R$ is too large.}.
The KK mass term, $in / R \psi _I ^{(n)} \chi _I ^{(n)}$,
comes from the $5D$ kinetic term.
Thus, we see that the 
$4D$ neutrino Yukawa couplings are suppressed by the
volume of the extra dimensions or, in other words, by the ratio $M_{\ast}/ 
M_{Pl}$.
When $h$ acquires a vev, we get Dirac mass terms for $\nu_f$ with
$\psi^{(n)}_I$ and $\chi^{(n)}_I$, denoted
by $m_{fI} \sim \lambda_{fI} v M_{\ast}/ M_{Pl}$ and $m^c_{fI} 
\sim \lambda ^c_{fI} 
v M_{\ast}/ M_{Pl}$, respectively ($v \approx 250$ GeV is the Higgs
vev). This analysis goes through for
any number of extra dimensions.

Consider the ``minimal'' model with $\mu _I = 0$ and $\lambda ^c_{fI} = 0$
and, to begin with, assume one SM neutrino and one $N$.
If $m R \ll 1$, then, to a good approximation,
$\nu$, $\psi ^{(0)}$ and $\chi ^{(n)}$, $\psi ^{(n)}$ $(n \neq 0)$
form Dirac fermions with masses $m$ and $n/R$ respectively, with
mixing between $\chi^{(n)}$ and $\nu$ given by $\sim mR / n \ll 1$.
$\chi^{(0)}$ decouples and is massless. 

For the case of three SM neutrinos $\nu_f$, 
we can introduce $3$ singlets, $N_I$
\footnote{The number of $2$-component Weyl spinors in
a Dirac spinor increases with the number of dimensions. For example,
in $6D$, it is $4$. Thus, in larger number of extra dimensions,
it might be possible to give all $3$ SM neutrinos Dirac masses using
only $1$ or $2$ singlets.}
so that the neutrino Dirac mass matrix is (up to small corrections
from $\nu_f$-$\chi^{(n)}_I$ mixing \cite{aetal, ddg})
\begin{equation}
\left( m_ {\nu} \right) _{fI} 
\sim m_{fI} \sim \lambda _{fI} v \frac{M_{\ast}}{M_{Pl}}
\sim \lambda _{fI} \left( \frac{M_{\ast}}{\hbox{TeV}} 
\right) 10^{-4} \; \hbox{eV}.
\label{mnu1}
\end{equation}

The $\nu - \chi^{(n)}$ mixing can have significant effects
on weak decays to $l \bar{\nu}$ as follows. The
SM neutrino
(weak eigenstate) is dominantly the lightest neutrino 
(with mass $m_{\nu}$) with small mixture
of heavier neutrinos (with mass $\sim n / R$):
\begin{equation}
\nu \approx \frac{1}{N} \left( \nu ^{(0)} _L + \sum _{n \neq 0}
\frac{mR}{n} \nu ^{(n)} _L
\right),
\end{equation}
where
$\nu^{(n)}$ are the mass eigenstates.
The ``normalization factor'' is 
\begin{equation}
N^2 \approx 1 + \sum _n \frac{ (m R)^2 }{n^2}  
\approx 1 + 
\frac{m^2}{M_{\ast}^2}
\frac{M_{Pl}^2}{M_{\ast}^2}  \approx 1 + 
\frac{\lambda ^2 v^2}{M_{\ast}^2},
\label{norm}
\end{equation}
where we have 
truncated the KK sum at $n / R \sim
M_{\ast}$, neglecting an $O(1)$ factor
in the summation.
For $\delta =2$, the sum is log-divergent and we get an additional
factor of $\sim \ln \left( M_{Pl} / M_{\ast} \right)$ in the sum. 
Thus,
the decay width to $\nu ^{(0)}$ is modified compared to SM
since $N^2 \neq 1$. Whereas
decays to $\nu ^{(n)}$'s $(n \neq 0)$ (if kinematically allowed)
are suppressed by small mixing ($\sim mR /n$) and have
a different phase space.

For example, consider the decays
$\pi^- \rightarrow e \bar{\nu}, \; \mu \bar{\nu}$
\cite{dk}. In the SM,
$\Gamma \left( \pi^- \rightarrow e \bar{\nu} \right)$
is suppressed by $m_e^2$ due to chirality flip. 
In the extra dimensional scenario, the chirality flip can occur
on the neutrino instead. In other words,
$\pi^-$
decays into $e_L$ and $\nu ^{(n)}  _R \sim
\psi ^{(n)}$, i.e., the heavier KK states, through the $m \nu \psi ^{(n)}$
term.
The large number of KK states up to $m_{\pi} - m_e$ enhance the 
effect,
whereas the same effect (relative to the SM)
in the case of
$\pi^- \rightarrow \mu _L
\psi^{(n)}$ is smaller.
For $\delta = 2$, this effect on
$\Gamma \left( \pi^- \rightarrow e \bar{\nu} \right)$ gives
a lower limit on $M_{\ast}$
of $O(1000)$ TeV for $\lambda \sim O(1)$ \cite{dk}.
Of course, $\lambda \sim O(1)$ with $M_{\ast} \sim 1000$ TeV gives
$m_{\nu _e} \sim 0.1$ eV, which might be too large.
The $\pi ^- \rightarrow e_L \psi ^{(n)}$ decay width scales
as $| \lambda |^2 \left( m_{\pi} / M_{\ast} \right) ^{\delta}$ so that
for $\delta \geq 3$, this effect is smaller than the effect of $N^2$ 
on
$\pi ^-$ decays into $e_R, \; \mu _R$ 
and $\nu ^{(0)} _L \sim \nu$. This
modifies the ratio of decay widths to $e$, $\mu$ since
the $m$'s and hence the normalization factors 
are different for $\nu _e$ and $\nu _{\mu}$; the lower limit on
$M_{\ast}$ is 
$O(10)$ TeV again for $| \lambda ^2_{\mu} 
- \lambda ^2_e | \sim O(1)$ \cite{dk}. 
We do get the
mass scale 
$\sim 10^{-5}$ (eV)$^2$
required for a solution to the solar neutrino anomaly
via matter-induced oscillations 
\cite{superk} for 
$M_{\ast} \sim O(10)$ TeV and $\lambda \sim O(1)$.

Coherent
conversion of SM neutrinos to singlet neutrinos
(due to the above mixing) in a supernova (SN) results in
energy loss, reducing its active neutrino flux. 
Since the SM neutrinos (unlike the singlet neutrino)
have weak interactions with the matter in the SN core, these
oscillations are enhanced by the MSW effect. 
These resonant oscillations are possible only if the mass
of the sterile neutrino state is not larger than
$\sim
\sqrt{E V} \sim 10$ keV, 
where $E \sim 100$ MeV is the neutrino energy in a SN and $V
\sim 10$ eV is
the potential in its core \cite{bcs}.
The survival probability of SM neutrino can be approximated
by the product of survival probabilities in each resonance ``crossed''
by the SM neutrino as it travels out of the SN core, $P_{\nu \nu} \approx
\prod _n P_n$ \cite{ds, bcs}. $P_n$ is approximately
independent of the mass of the $n^{\hbox{th}}$
resonance and is given by \cite{ds, bcs} 
\begin{equation}
P_n \approx \exp \left( - \frac{\pi}{2}
\frac{ 4 m^2 r_{\hbox{core}} }{E}\right) \sim \exp \left( -
\frac{m^2}{10^{-3} \; (\hbox{eV}) ^2} \right),
\label{pn}
\end{equation}
where $r_{\hbox{core}} \sim 10$ km is the radius of the SN core.
To explain the atmospheric neutrino anomaly via oscillations, 
we require $m^2 \; (\hbox{for} \; \nu _{\mu} \; \hbox{or}
\; \nu _{\tau}) \;  \sim
\Delta m^2 _{\hbox{atm}} \sim 10^{-3}$ (eV)$^2$ \cite{superk}
so that
each $P_n$ is $\sim 1/3$. 
Thus, even if only one resonance is
crossed, the energy loss from a SN due to sterile neutrinos will
be comparable to that due to active neutrinos.
The agreement of the measured duration and number of neutrino events 
from SN$1987$a with
the prediction of SN models (taking into account theoretical uncertainties)
indicates
that a new channel for energy loss from the SN 
should be less effective than the standard neutrino channel; otherwise
the neutrino signal duration will be halved, which is not allowed
\cite{raffelt}. 
Therefore, in this case,
the measurement of SN$1987$a neutrino flux 
implies that no resonance can be crossed so that 
the
mass of the lightest KK state, which is
$1/ R$, should be larger than
$\sim 
10$ keV
\cite{bcs}.
This implies that the KK states 
are much heavier than SM neutrinos and hence
both the solar and atmospheric neutrino anomalies have to be
explained by oscillations
among active neutrinos,
governed by the mass matrix $m_{\nu}$ above.

Loop diagrams involving KK neutrino tower and
longitudinal $W$ 
contribute to $\mu \rightarrow e \gamma$ \cite{fp, ip}
and $(g - 2) _{\mu}$ \cite{graesser, ng}.
The coefficient of the dimension-$5$ operator relevant for these two
processes,
$F^{\mu \nu} \bar{l} \sigma _{\mu \nu} l$
($l = e, \mu, \tau$), generated 
by the KK neutrino exchange, is approximately
\begin{equation}
e \; \frac{m_l}{8 \pi^2 v^2} 
m m^{\dagger} \sum _n \frac{R^2}{n^2} \sim
e \; \frac{m_l}{8 \pi^2 v^2}
\frac{m m^{\dagger}}{M_{\ast}^2} \frac{M_{Pl}^2}{M_{\ast}^2}.
\end{equation}
The amplitude is enhanced as compared
to $4D$ case by the large number of KK states
(we have truncated the KK sum at $M_{\ast}$).
Since the neutrino masses are given by $m$ (see Eq. (\ref{mnu1})),
it is obvious that (as in $4D$)
there is a direct correlation 
between neutrino 
oscillations
and the contributions to
$\mu \rightarrow e \gamma$ from these loop diagrams. 
For example, in the two flavor case, we get
\begin{eqnarray}
{\cal A} \left( \mu \rightarrow e \gamma \right) & \propto &
\frac{ \left( m m^{\dagger} \right) _ 
{e \mu}}{M_{\ast}^2} \; \frac{M_{Pl}^2}{M_{\ast}^2} \nonumber \\
 & \sim & \frac{ | m^2 _{\nu _1} - m^2 _{\nu _2} | }{M_{\ast}^2}
\sin 2 \theta 
\frac{M_{Pl}^2}{M_{\ast}^2},
\label{mueg1}
\end{eqnarray}
where $m_{\nu _1}$, $m_{\nu_2}$ and $\theta$
are the masses and mixing angle of the two neutrinos
obtained from Eq. (\ref{mnu1}).
This was used in \cite{fp}
to obtain lower limits on $M_{\ast}$ of 
$O(10-100)$ TeV for
$\theta \sim \pi /4$ and $| m^2 _{\nu _1} - m^2 _{\nu _2} |
\sim 10^{-5}$ (eV)$^2$ (as 
relevant for large mixing angle 
solar oscillations). In the case of three flavors with $2-3$ mixing
angle $\phi$, but
no $1-3$ mixing, the above expression
is simply multiplied by $\cos \phi$.

In the minimal model, we see from Eq. (\ref{mnu1})
that to get
$m_{\nu}^2 \sim \Delta m_{\hbox{atm}}^2$ we require $M_{\ast}
\stackrel{>}{\sim} 100$ TeV
if $\lambda \sim O(1)$. Such high values of $M_{\ast}$
are disfavored by the motivation to 
solve the hierarchy problem \cite{add}.
Of course, we can choose
$\lambda \gg 1$ and obtain $m^2 _{\nu} \sim
\Delta m_{\hbox{atm}}^2$
for $M_{\ast} {\sim} 10$ TeV \cite{us}
\footnote{For $M_{\ast}
\stackrel{<}{\sim} 10$ TeV, it is not possible to obtain
$\Delta m_{\hbox{atm}}^2$ even for $\lambda \gg 1$
since, due to the normalization factor, there is an upper
limit on $m_{\nu}$, for a given $M_{\ast}$
\cite{aetal, us}.},
but then the $(4 + \delta)
D$ theory
might reach strong coupling. 
In other words,
from
the $4D$ point of view, the Yukawa coupling 
$\sim \lambda M_{\ast} / M_{Pl}
\sim m_{\nu} / v$ is very small at tree level (or at low energies)
even though $\lambda \gg 1$ in this case.
But, since the $4D$ coupling
``runs'' with power of energy due to the multiplicity of KK states,
it might reach its Landau pole near $M_{\ast}$.
For this reason, we will consider
$\lambda \stackrel{<}{\sim} O(1)$ throughout this
paper.

With the motivation
of obtaining
the neutrino mass scale $\Delta m_{\hbox{atm}}^2$ with $M_{\ast} \sim$ TeV
and $\lambda \sim O(1)$, we now study
two modifications
of the minimal model. 
In the first model, the singlet neutrino
propagates in a sub-space of the full extra dimensional space
where gravity propagates.
In the second model, we consider the effect of non-zero
Dirac masses for the singlets and of lepton number violating couplings,
$\lambda ^c$.
We will also keep an eye on the correlation between
neutrino masses and contribution to
$\mu \rightarrow e \gamma$ and the effect on
$\pi^- \rightarrow e \bar{\nu}, \; \mu \bar{\nu}$
in these models.

{\bf \large 2) Sub-space}

First, consider the case where
singlet neutrino propagates in 
$\delta _{\nu} < \delta$ dimensions
\cite{aetal}. Assuming that
all extra dimensions are of size $R$, in this case, we get
the neutrino Dirac mass matrix
\begin{equation}
\left( m_{\nu} \right) _{fI} \sim m_{fI} \sim
\frac{ \lambda _{fI} } 
{ \sqrt{ \left(R M_{\ast} \right)^{\delta _{\nu}} } } v
\sim \lambda _{fI} v \left( \frac{M_{\ast}}{M_{Pl}}
\right) ^{\delta_{\nu} / \delta} . 
\label{mnu2}
\end{equation}
Thus, for  
$\delta _{\nu} =5$ and $\delta =6$ and with
$M_{\ast} \sim$ TeV, $\lambda \sim O(1)$, we get $m ^2 \sim
\Delta 
m_{\hbox{atm}}^2$ \cite{aetal}.
To obtain the neutrino
mass scale required for a solution to the solar neutrino anomaly
via matter-induced oscillations,
$\Delta m^2 _{\hbox{sol}} \sim 10^{-5}$ (eV)$^2$, we can choose 
the corresponding $\lambda \sim O(0.1)$. 

In this case, the dimension-$5$ operator relevant for
$\mu \rightarrow e \gamma$ has the coefficient  
\begin{eqnarray}
e \; \frac{m_{\mu}}{8 \pi^2 v^2} \frac{ \left(
m m^{\dagger} \right) _{e \mu}}
{M_{\ast}^2}
\sum \frac{R^2}{n^2} & \sim & 
e \; \frac{m_{\mu}}{8 \pi^2 v^2} \frac{
\left( m m^{\dagger} \right) _{e \mu}}{M_{\ast}^2}
\left( \frac{M_{Pl}^2}{M_{\ast}^2} \right) ^{\delta_{\nu} / \delta}
\nonumber \\
 & \propto & \frac{ | m^2 _{\nu _1} - m^2_{\nu_2} | }{M_{\ast}^2}
\sin 2 \theta
\left( \frac{M_{Pl}^2}{M_{\ast}^2} \right) ^{\delta_{\nu} / \delta},
\end{eqnarray}
where $m_{\nu _1}$, $m _{\nu _2}$ and $\theta$ are 
the neutrino masses and mixing angle as obtained from
Eq. (\ref{mnu2}). 
We see that this contribution is 
suppressed compared to $\delta _{\nu} = \delta$ (for the {\em same} values of
$m$ and $M_{\ast}$) due to
smaller number of KK states. 
Thus, the lower limits on $M_{\ast}$ of $O(10-100)$ TeV
obtained for the minimal model
\cite{fp} (for
$\theta \sim \pi / 4$ and
$| m^2 _{\nu _1} - m^2 _{\nu _2} | \sim 10^{-5}$ (eV)$^2$)
can be relaxed by a factor of
$O(20)$, assuming
$\delta _{\nu} =5$ and $\delta =6$. 
The lower limit now becomes $M_{\ast} \sim$ few TeV.

In terms of $\lambda$ (instead of $m$), the coefficient
of the dimension-$5$ operator is
\begin{equation}
e \; \frac{m_{\mu}}{8 \pi^2} \frac{ \left(
\lambda \lambda ^{\dagger} \right) _{e \mu}}
{M_{\ast}^2},
\end{equation}
i.e., for {\em fixed} $\lambda$,
it is independent of $\delta _{\nu}$ or $ \delta$
\cite{ip}.
But, with $\delta _{\nu} =5$, $\delta =6$ and $M_{\ast} \sim$ few TeV,
we require $\left( \lambda \lambda ^{\dagger} \right)
_{e \mu} \sim O(10^{-2})$
to obtain the mass scale for solar neutrino oscillations
(as shown above). Whereas with $\delta _{\nu} = \delta$
and for $M_{\ast} \sim O(10-100)$ TeV,
we require $\left( \lambda \lambda ^{\dagger} 
\right) _{e \mu} \sim O(1)$.
Hence the above coefficient is the same for 
these two parameter sets
(which give the same neutrino masses), 
in agreement with the analysis in terms of $m$. 

Reference \cite{ip} also considers the constraints from
$\mu \rightarrow 3e$ and $\mu \rightarrow e$ conversion in nuclei.
For these processes, loop contribution due to KK neutrino tower to
the effective $Z$-$\mu$-$e$ coupling
(in addition to the
$\gamma$-$\mu$-$e$ coupling) has to be included.
This coupling
depends on $\sim \lambda ^4 v^2 / M_{\ast}^2$ \cite{ip}
(dropping the flavor indices on $\lambda$ for simplicity), 
unlike the
$\gamma$-$\mu$-$e$ coupling 
which depends on $\sim \lambda ^2 v^2 / M_{\ast}^2$
as above. 
The experimental bounds on 
$\mu \rightarrow 3e$ and $\mu \rightarrow e$ conversion in nuclei
give the lower limit $M_{\ast} / \lambda ^2 \sim 200-300$ TeV \cite{ip}
which, for $\lambda \sim 1$, is stronger than that from
$\mu \rightarrow e \gamma$.
However, in this sub-space case,
as mentioned above, to get $m_{\nu}^2 \sim 
\Delta m^2 _{\hbox{sol}}$ with
$M_{\ast} \sim$ few TeV, we require $\lambda \sim O(0.1)$
and hence this quantum gravity scale
is consistent with
$\mu \rightarrow 3e$ and $\mu \rightarrow e$ conversion in nuclei.
  
In this sub-space scenario, 
the effect on $\pi^-
\rightarrow 
e \bar{\nu}, \; \mu \bar{\nu}$ decays is due to the normalization factor
$N^2$ 
(since $\delta _{\nu} =5$) and thus 
also depends on $\sum _n (mR)^2 / n^2$ (see Eq. (\ref{norm})). As above,
this factor 
is smaller than in the minimal
model
for the same values of $m$ and $M_{\ast}$. Therefore,
the lower limit on $M_{\ast}$ from $\pi^- \rightarrow e \bar{\nu}, \; 
\mu \bar{\nu}$
decays is also reduced from $O(10)$ TeV
(obtained for the minimal model
\cite{dk}, assuming $\lambda \sim O(1)$ which gives
$m_{\nu}^2 \sim 10^{-5}$ (eV)$^2$)
to $\sim$ few TeV.

With $\delta =6$, we get $1/ R \sim O(10 - 100)$ MeV so that
the SN$1987$a constraint is satisfied.

{\bf \large 3) See-saw}

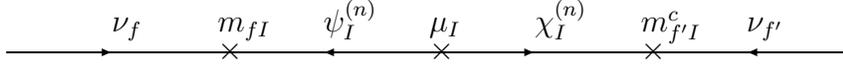
\begin{figure}
\setlength{\unitlength}{0.8pt}
\begin{picture}(500, 100)(25,0)
\put(50,50){\vector(1,0){50}}
\put(100,60){$\nu _f$}
\put(100,50){\line(1,0){100}}
\put(150,60){$m_{fI}$}
\put(150,47){$\times$}
\put(250,60){$\mu _I$}
\put(250,47){$\times$}
\put(200,60){$\psi _I ^{(n)}$}
\put(250,50){\vector(-1,0){50}}
\put(250,50){\vector(1,0){50}}
\put(300,50){\line(1,0){100}}
\put(300,60){$\chi _I ^{(n)}$}
\put(350,47){$\times$}
\put(350,60){$m^c _{f^{\prime} I}$}
\put(450,50){\vector(-1,0){50}}
\put(400,60){$\nu _{f ^{\prime}}$}
\end{picture}
\caption{See-saw mechanism for generating SM neutrino masses in the case
$\mu _I \neq 0$ and $m^c \neq 0$.
There is another diagram obtained by
$m \leftrightarrow m^c, \; \psi _I ^{(n)} \leftrightarrow \chi _I ^{(n)}$. }
\label{mnu}
\end{figure}

Next, consider the case with non-zero Dirac mass $\mu_I (\ll
M_{\ast})$ for the singlets \cite{ip, lukas1, lukas2}. 
In the limit $\mu \gg m$ and
$m^c \approx 0$, 
the linear combinations
$\sim \left( 
\chi _I^{(n)} + m_{fI} / \sqrt{\mu_I^2 + n^2 / R^2} \; \nu_f
\right)$ form Dirac pairs
with $\psi _I^{(n)}$'s of masses $\sim
\sqrt{\mu^2_I + n^2 / R^2}$.
Of course, with $\lambda ^c \approx 0$, 
there are $3$ massless neutrinos
which are dominantly $\nu_f$'s, even though a priori
all fermions have mass terms. The reason is that in this case
we can
define a conserved lepton number with charges $+1$ for $\nu$ and
$\chi$ and $-1$ for $\psi$ so that only Dirac masses are allowed. Thus,
there are $3$ ``unpaired'' fermions
with charge $+1$ which are the $\nu _f$'s. In the minimal model
($\mu =0$), these massless fermions are
$\chi_I^{(0)}$'s as mentioned earlier.
We require lepton number violation, for example,
$m^c \neq 0$, so that these massless neutrinos can get
Majorana 
masses
$\nu _f \nu _{f^{\prime}}$ \cite{lukas1,
lukas2}. Then, 
the see-saw mechanism
(see Fig. \ref{mnu}) gives 
\begin{equation}
\left( m_{\nu} \right) _{f f^{\prime}} \sim
\sum _I \sum _n \frac{m_{f I} \mu_I 
m^{c} _{f^{\prime} I} +
m^{c}_{f I} \mu_I m _{f^{\prime} I} }{\mu_I^2 + n^2 / R^2}.
\end{equation}
In this case, one singlet $N$ suffices to give masses to all $3$
SM neutrinos, but in general, one can have many singlets with
different 
$\mu _I$'s.

In the case of $\delta \geq 2$,
the see-saw is ``divergent''due to $\sum _n 1/ n^2$. If
we truncate the KK sum at $M_{\ast}$, then we get the
neutrino Majorana mass matrix
\begin{eqnarray}
\left( m_{\nu} \right) _{f f^{\prime}} & \sim &
\sum _I \left( m_{fI} \mu_I m^c_{f^{\prime} I} +
m^c _{fI} \mu _I
m _{f^{\prime} I} \right) R^{\delta} M_{\ast}^{\delta -2} \nonumber \\
 &
\sim & \left( m \mu m^{c \; T}
+ m^c \mu m^T \right) _{f f^{\prime}}
\frac{M_{Pl}^2}{M_{\ast}^4} \nonumber \\
 & \sim & \left( \lambda \mu \lambda ^{c \; T}
+ \lambda ^c \mu \lambda ^T \right)_{f f^{\prime}} \frac{v^2}{M_{\ast}^2}.
\label{mnu3}
\end{eqnarray}
The KK sum is log-divergent for $\delta = 2$ so that the above
result is multiplied by a factor 
of $\sim 
\ln \Big[ M^2_{\ast} / \left( \mu ^2 + 1 / R^2
\right) \Big] \sim O(10)$ and was mentioned
in \cite{lukas1}. 
We see from Eq. (\ref{mnu3}) that for
$M_{\ast} \sim$ TeV, $\delta
\geq 2$ and $\lambda$, $\lambda ^c \sim O(1)$,
we get $m_{\nu}^2 \sim \Delta m_{\hbox{atm}}^2$
by choosing
$\mu \sim $ eV. If all singlets have Dirac masses of $O(
\hbox{eV})$, then
to get $m_{\nu}^2 \sim \Delta m^2 _{\hbox{sol}}$, we can choose
the corresponding $\lambda$, $\lambda ^c \sim O(0.1)$.

We next consider the constraints from SN$1987$a on this model.
The analysis is different than in 
the minimal model since in this case $m _{\nu} \not \! \sim m$
(the latter governs the mixing between SM and sterile neutrinos
and hence the energy loss in SN$1987$a)
and also the
masses of the sterile neutrinos are given by $\sim
\sqrt{ \mu ^2 + n^2 / R^2 }$ instead of $n/R$.
Thus, one way to evade the SN$1987$a constraint, independent of
the number and 
the size of the extra dimension, is to choose
$\mu _I \stackrel{>}{\sim} 10$ keV so that no
resonance can be crossed in the SN$1987$a core.
Then, the solar and atmospheric neutrino oscillations
have to involve only active neutrinos.
With $\mu \sim 10$ keV,
we require
$\lambda \lambda ^c \sim O(10^{-4})$,
$O(10^{-5})$ to get $m_{\nu}^2 \sim
\Delta m^2 _{\hbox{atm}}$,
$\Delta m^2 _{\hbox{sol}}$, respectively, i.e,
$\lambda$, $\lambda ^c \sim O(1)$ will give too large
$m_{\nu}$ (see Eq. (\ref{mnu3})).

If $\delta \geq 4$, then we get $1/ R > 10$ keV so that
again no resonance is crossed and 
the SN$1987a$ constraint is satisfied for any value of $\mu$.

The remaining cases are $\mu _I \ll
10$ keV and $\delta = 2,3$.
The number of resonances crossed,
i.e., the number of sterile neutrino states lighter than $10$
keV, is $n_{\hbox{res}} \sim ( R \; 10 \; \hbox{keV} )^{\delta}$, and
we have to consider each value of $\delta$ separately.

For $\delta = 2$ and $M_{\ast} \sim 1-10$ TeV, we get 
$1/R \sim 0.01$ eV so that $n_{\hbox{res}} \sim 10^{12}$.
As mentioned earlier, 
we require $P_{\nu \nu} \approx \left( P_n
\right) ^{n_{\hbox{\tiny res}}} \stackrel{>}{\sim}
1/2$ so that the SN energy loss due to sterile neutrinos is
less than that due to active neutrinos.
Thus, we get the constraint
$m^2$, $m^{c \; 2} \stackrel{<}{\sim} 10^{-15}$ (eV)$^2$. 
Then, the 
active neutrino
masses are too small to account for solar and atmospheric
neutrino oscillations. 
The $\nu _e - \chi^{(n)} / \psi ^{(n)}$ mixing is also
too small to be relevant for explaining the
solar neutrino anomaly by oscillations of $\nu _e$ to sterile
neutrinos.
Thus, $\mu \ll 10$ keV is ruled out by SN$1987$a constraint.

For $\delta =3$ and $M_{\ast} \sim 1-10$ TeV, we get $1/ R \sim
O(100 \; \hbox{eV} - 1
\; \hbox{keV})$ so that $n _{\hbox{res}} \sim 10^{3} - 10^{6}$
\footnote{$R$ and hence $n_{\hbox{res}}$ is very sensitive
to the value of $M_{\ast}$.}. Hence, 
the SN$1987$a constraint, $P_{\nu \nu} \sim 1$, requires
$m^2$, $m^{c \; 2} \stackrel{<}{\sim} 10^{-6} - 10^{-9}$ (eV)$^2$, i.e.,
$\lambda, \; \lambda ^c \sim O(1)$ might be  
allowed
(depending on $M_{\ast}$).
Thus, 
we can choose $\mu \sim$ eV
to get $m_{\nu}^2 \sim \Delta m^2 _{\hbox{atm}}$
for $M_{\ast} \sim$ TeV and $\lambda, \; \lambda ^c \sim O(1)$
while (marginally) satisfying the SN$1987$a constraint.

Of course, for $\delta \geq 3$ and for any value of $\mu$,
the sterile neutrinos are heavier than
$1/R \stackrel{>}{\sim} 100$ eV so that they cannot be 
directly involved in
oscillations of SM neutrinos in the sun or in the atmosphere.

The coefficient of the dimension-$5$ operator,
$F^{\mu \nu} \bar{l} \sigma _{\mu \nu} l$, generated 
by exchange of KK neutrinos, is similar
to the earlier case, with an additional contribution from
$m^c$:
\begin{equation}
e \; \frac{m_l}{8 \pi^2 v^2} \frac{m m^{\dagger} + m^c m^{c \; \dagger}}
{M_{\ast}^2} \frac{M_{Pl}^2}{M_{\ast}^2}.
\label{mueg3}
\end{equation}
Thus, we see that both $(g - 2) _{\mu}$ and
BR$\left( \mu \rightarrow e \gamma \right)$ depend on 
$\left( m m^{\dagger} + m^{c} m^{c \; \dagger}
\right)
$ whereas $m_{\nu}$ depends on $\left( m \; \mu \; m^{c \; T} +
m^{c} \; \mu \; m^T \right)$. Therefore, it is clear that the
parameters for neutrino
oscillations and for
$\mu \rightarrow e \gamma$,
$(g - 2) _{\mu}$ may not be related in this class of models. In
particular
\begin{equation}
{\cal A} \left( \mu \rightarrow e \gamma \right) 
\propto \left( m m^{\dagger} + m^{c} m^{c \; \dagger}
\right)
_{e \mu} \not \! 
\sim | m^2 _{\nu _1} - m^2 _{\nu_2} | \sin 2 \theta,
\end{equation}
where the masses $m _{\nu _1}$, $m_{\nu_2}$ and mixing angle
$\theta$ 
(relevant for $\nu_e$-$\nu _{\mu}$ oscillations) are obtained from
$m_{\nu}$ in Eq. (\ref{mnu3}).

To illustrate the above point, 
consider
for
simplicity only $\nu_e$ and $\nu_{\mu}$ and 
suppose
we have 
$2$ singlets with the same $\mu$'s. Assume 
\begin{eqnarray}
m & \propto & \left( \begin{array}{cc}
0 & 0 \\
\rho & 1 
\end{array}
\right), \\ 
m^c & \propto & \left( \begin{array}{cc}
1 & 0 \\
0 & \sigma 
\end{array} 
\right)
\end{eqnarray}
with $\rho$, $\sigma \sim O(1)$. In this example, 
$\mu \rightarrow e \gamma$ depends on 
\begin{equation}
\left( \begin{array}{cc}
1 & 0 \\
0 &  1 + \sigma^2 + \rho^2 
\end{array}
\right),
\end{equation}
whereas 
\begin{equation}
m_{\nu} \propto
\left( \begin{array}{cc}
0 & \rho \\
\rho & 2 \sigma 
\end{array}
\right)
\end{equation}
so that $\nu_{\mu} - \nu_e$ mixing can be large (with non-degenerate 
neutrinos)
as required for solar oscillations with large mixing angle,
but the loop contribution to
$\mu \rightarrow e \gamma$ is zero.

Even if the flavor structures of $m$ and $m^c$ are similar, 
BR$\left( \mu \rightarrow e \gamma \right)$ might not be correlated
with neutrino masses since,
for given $M_{\ast}$,
$\mu \rightarrow e \gamma$ depends only on $\lambda$ and $\lambda ^c$,
whereas $m_{\nu}$ depends also on $\mu$.
For $\delta =2$, as explained above, to satisfy the constraints
from SN$1987$a, we might have to choose
$\mu \sim 10$ keV and hence $\lambda \lambda ^c \sim 10^{-5}$
to get $m_{\nu}^2 \sim \Delta m^2 _{\hbox{sol}}$.
Then, 
$M_{\ast} \sim$ few TeV is consistent with the limit on 
BR$\left( \mu \rightarrow e \gamma \right)$.
For $\delta \geq 3$, $\mu$ is not constrained by SN$1987$a.
So, we can choose
the relevant $\lambda$, $\lambda ^c$
small enough such that $M_{\ast} \sim$ few TeV is
consistent with $\mu \rightarrow e \gamma$  
and, at the same time, we can get
the required $m_{\nu _e}$, $m_{\nu _{\mu}}$ by choosing
the corresponding $\mu _I$'s appropriately.
As mentioned in section $2$, loop contributions
to $\mu \rightarrow
3e$ and $\mu \rightarrow e$ conversion in nuclei also depend on
$\lambda$ and $\lambda ^c$ and hence are suppressed
if $\lambda$, $\lambda ^c$ are small.

A similar analysis shows that $M_{\ast} \sim$ few TeV can
be consistent with $\pi ^-
\rightarrow e \bar{\nu}, \; \mu
\bar{\nu}$ since,
for given $M_{\ast}$ and for $\mu \ll m_{\pi}$, 
the effect on $\pi ^-$ decays also depends only on
$\lambda$ and $\lambda ^c$.

{\bf \large 4) Conclusion}

In summary, we have studied two ``non-minimal'' models
with singlet neutrino in large extra dimensions and TeV scale quantum gravity.
These models can accommodate the mass scale relevant for
atmospheric neutrino oscillations even with $O(1)$ 
higher-dimensional Yukawa couplings and
$M_{\ast} \sim $ a few TeV, unlike the minimal model. 
The first model has 
singlet neutrino propagating in smaller number of extra dimensions
as compared to gravity whereas the second model has small Dirac mass
terms for the singlets
and lepton number violating couplings. In both models, the constraints
on $M_{\ast}$ from BR$\left( \mu \rightarrow e \gamma \right)$ 
and $\pi \rightarrow e \bar{\nu}, \; \mu \bar{\nu}$
decays can be significantly weakened
as compared to the minimal model so
that $M_{\ast} \sim$ few TeV is consistent with
these decays. 
Also, due to the SN$1987$a constraint,
active-sterile neutrino oscillations are strongly disfavored 
in both the minimal
and the modified models.

{\bf Acknowledgments}
This work is supported by DOE Grant DE-FG03-96ER40969.
K.A. thanks the Aspen Center for Physics for hospitality during 
part of this work.


\begin{thebibliography}{99}
\bibitem{add}N. Arkani-Hamed, S. Dimopoulos, G. Dvali,
hep-ph/9803315, Phys. Lett.
B 429 (1998) 263 and
I. Antoniadis et al, hep-ph/9804398, Phys. Lett. B 436 (1998) 257.
\bibitem{aetal}N. Arkani-Hamed et al., talk
by S. Dimopoulos at SUSY 98, Oxford,
July 11-17, 1998 and hep-ph/9811448.
\bibitem{ddg}K.R.
Dienes, E. Dudas, T. Gherghetta, hep-ph/9811428, Nucl. Phys.
B 557 (1999) 25.
\bibitem{others}
For further studies of these
models see references
\cite{bcs}-\cite{us}
below and
A. Pilaftsis, hep-ph/9906265, Phys. Rev. D 60 (1999) 105023;
R.N. Mohapatra, S. Nandi, A. Perez-Lorenzana,
hep-ph/9907520, Phys. Lett. B 466 (1999) 115;
G.C. McLaughlin, J.N. Ng, hep-ph/9909558, Phys. Lett.
B 470 (1999) 157 and nucl-th/0003023;
R.N. Mohapatra, A. Perez-Lorenzana, hep-ph/9910474, Nucl. Phys. B 576 (2000) 
466 and
hep-ph/0006278;
A. Ioannisian, J.W.F. Valle, hep-ph/9911349;
K.R. Dienes, hep-ph/0004129;
K.R. Dienes, I. Sarcevic, hep-ph/0008144.
\bibitem{bcs}R. Barbieri, P. Creminelli, A. Strumia, hep-ph/0002199.
\bibitem{ds}
G. Dvali, A. Smirnov, hep-ph/9904211, Nucl. Phys. B
563 (1999) 63.
\bibitem{fp}A.E. Faraggi, M. Pospelov, hep-ph/9901299, Phys. Lett. B
458 (1999) 237.
\bibitem{ip}A. Ioannisian, A. Pilaftsis, hep-ph/9907522, Phys. Rev.
D 62 (2000) 066001 and hep-ph/0010051.
\bibitem{graesser}M. Graesser, hep-ph/9902310, Phys. Rev. D 61 (2000) 074019.
\bibitem{ng}G.C. McLaughlin, J.N. Ng, hep-ph/0008209.
\bibitem{dk}A. Das, O.C.W. Kong, hep-ph/9907272, Phys. Lett. B 470 (1999)
149.
\bibitem{lukas1}A. Lukas, A. Romanino, hep-ph/0004130.
\bibitem{lukas2}A. Lukas et al., hep-ph/0008049.
\bibitem{us}
K. Agashe, N.G. Deshpande, G.-H. Wu, hep-ph/0006122, 
Phys. Lett.
B. 489 (2000) 367.
\bibitem{superk}
See, for example, talk by E.T. Kearns at ICHEP, July 27 - August 2, 2000,
Osaka, Japan, 
http://ichep2000.hep.sci.osaka-u.ac.jp/scan/0801/pl/kearns/index.html.
\bibitem{raffelt}G.G. Raffelt, {\it Stars as Laboratories for Fundamental
Physics}, University of Chicago Press (1996) and
Ann. Rev. Nucl. Part. Sci. 49 (1999) 163.

\end{thebibliography}
\end{document}